\documentclass[aps,prb,reprint,amssymb,amsmath,superscriptaddress]{revtex4-2}
\usepackage{graphicx}
\usepackage{epstopdf}

\usepackage[english]{babel}
\bibpunct{[}{]}{,}{n}{}{}
\begin{document}

\title{Population of spin–orbit-split electron branches in semimetallic HgTe quantum wells}

\author{G. M.~Minkov}

\affiliation{School of Natural Sciences and Mathematics, Ural Federal University,
620062 Ekaterinburg, Russia}

\author{O.\,E.~Rut}
\affiliation{School of Natural Sciences and Mathematics, Ural Federal University,
620062 Ekaterinburg, Russia}

\author{A.\,A.~Sherstobitov}

\affiliation{School of Natural Sciences and Mathematics, Ural Federal University,
620062 Ekaterinburg, Russia}

\affiliation{M.~N.~Miheev Institute of Metal Physics of Ural Branch of
Russian Academy of Sciences, 620137 Ekaterinburg, Russia}

\author{A.\,V.~Germanenko}
\affiliation{School of Natural Sciences and Mathematics, Ural Federal University,
620062 Ekaterinburg, Russia}

\author{S.\,A.~Dvoretski}

\affiliation{Institute of Semiconductor Physics RAS, 630090
Novosibirsk, Russia}

\author{N.\,N.~Mikhailov}

\affiliation{Institute of Semiconductor Physics RAS, 630090
Novosibirsk, Russia}
\affiliation{Department of Physics, Novosibirsk State University, Novosibirsk 630090, Russia}

\begin{abstract}
We present the results of experimental studies on magnetoresistance, the Hall effect, and Shubnikov–de Haas oscillations performed in gated heterostructures with HgTe-based gapless semiconductor quantum wells. We show that in quantum wells with widths corresponding to a semimetallic charge carrier spectrum, electrons -- in contrast to the idealized model—populate both spin–orbit -- split branches of the spectrum. This occurs regardless of whether the Fermi level lies exclusively in the conduction band or within the conduction–valence band overlap region.
\end{abstract}


\maketitle

\label{sec:intr}

\begin{figure}[h]
\includegraphics[width=\linewidth,clip=true]{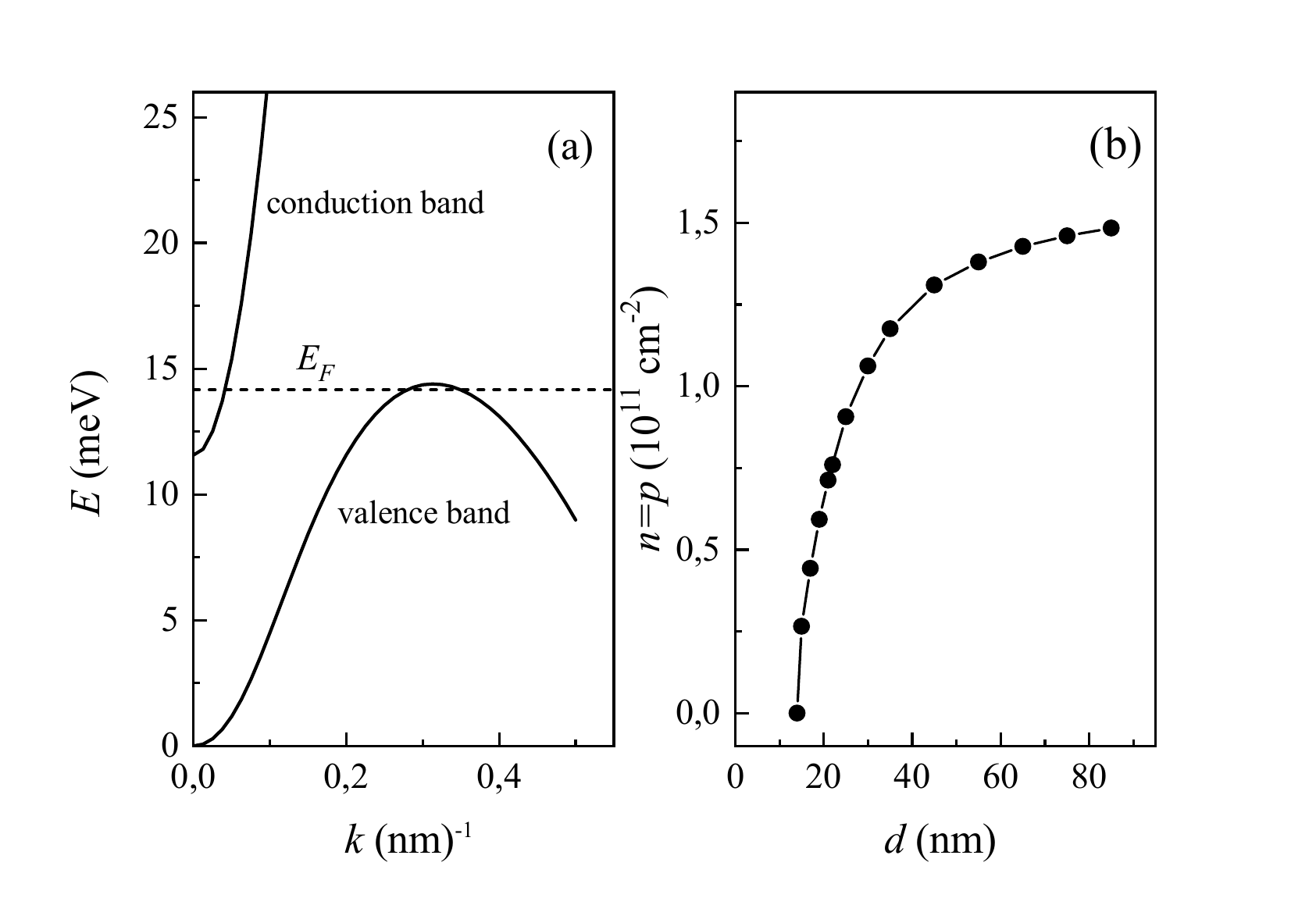}
\caption{(Color online) (a) The energy spectrum of the symmetric HgTe quantum well. (b) The $d$ dependence of electron density when the Fermi level is fixed on the top of the valence band.}
\label{f10}
\end{figure}

The energy spectrum of HgTe-based quantum wells (QWs) changes radically with increasing QW width ($d$). At $d < d_c\simeq (6.3 - 6.5)$~nm, it resembles that of a conventional semiconductor, evolving into a quasi-Dirac spectrum (linear at small wave vectors $k$) at the critical width $d =d_c)$~nm. Upon a further increase in $d$, the spectrum becomes inverted as the states forming the conduction and valence bands swap places. Finally, at $d > 14-15$~nm, the conduction and valence bands overlap, giving rise to a semimetallic state where electrons and holes coexist over a certain range of Fermi energies.  As an example, Fig.~\ref{f10}(a) shows the spectrum of the conduction band and the valence band of an undoped HgTe quantum well with $d = 32$~nm, and Fig.~\ref{f10}(b) shows the dependence of the electron density (in the absence of doping, the densities of electrons and holes are equal) on the width of the quantum well. As seen the overlap of the conduction and the valence bands (the transition to a semimetallic spectrum) and the appearance of electrons against the background of the valence band occur at $d \simeq 14$~nm. The electron density in intrinsic QWs increases rapidly up to $d \simeq (10–40)$~nm and grows significantly more slowly with further $d$ increase.

\begin{figure}[h]
\includegraphics[width=\linewidth,clip=true]{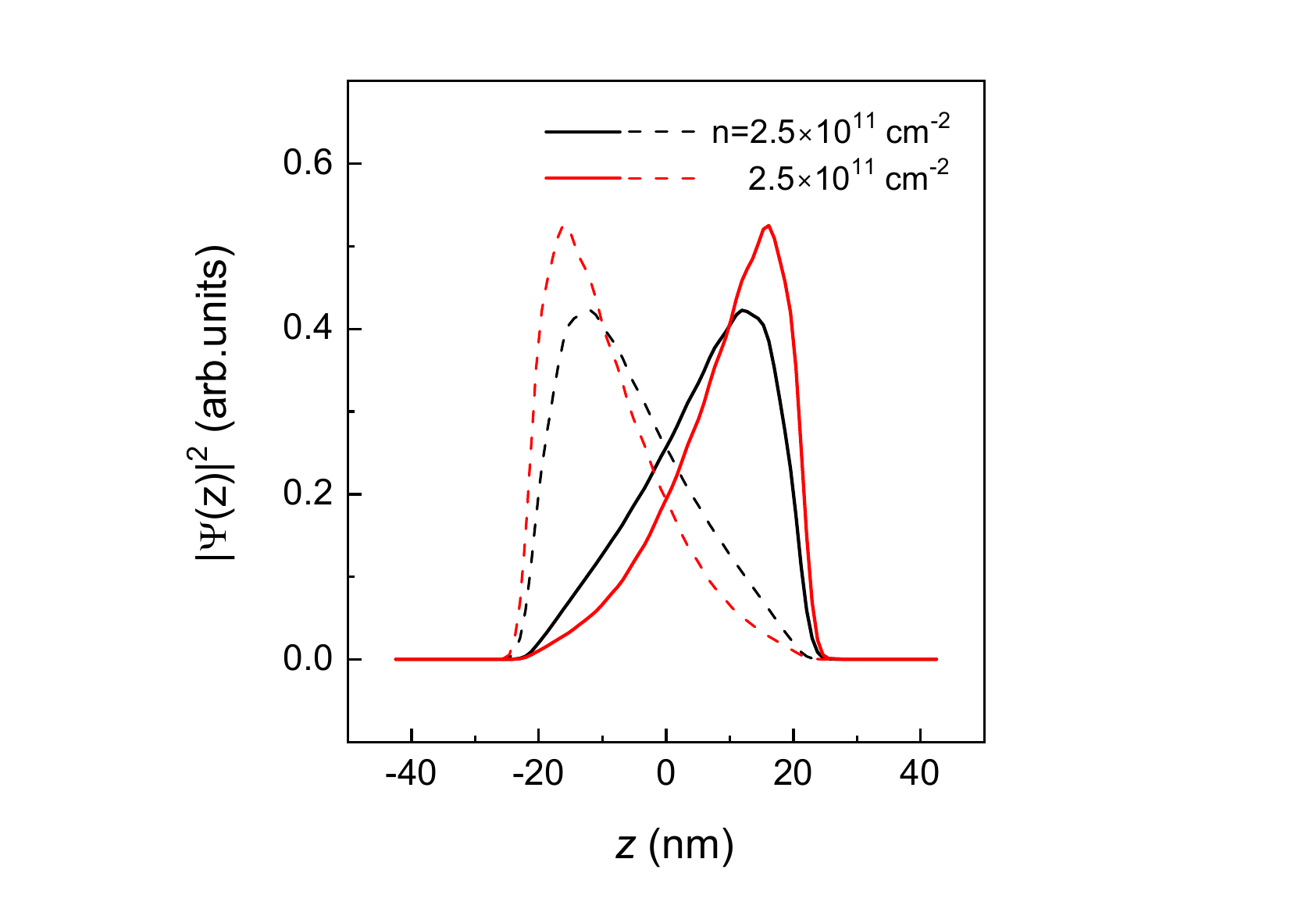}
\caption{(Color online) The distribution of electron density in the direction perpendicular to the QW plane at an energy equal to the Fermi energy for two electron densities. The solid and dashed lines correspond to two ``spin'' states that differ in the sign of $k$.}
\label{f20}
\end{figure}

The specific features of the spectrum of HgTe-based heterostructures, which are associated with strong spin–orbit interaction, lead to the fact that the wave functions of various ``spin'' states with  $k \neq 0$ are shifted along the $z$‑axis, where $z$ is the normal to the plane of the quantum well. This shift increases with the magnitude of the quasi‑momentum and depends on the width of the quantum well. For example, Fig.~\ref{f20} shows the results of a self-consistent calculation for the electron density distribution along the $z$-axis performed within the framework of the $kP$-model \cite{Minkov22} for the same structure at two values of $k$, corresponding to electron densities
$2.5\times 10^{11}$~cm$^{-2}$ and $6.0\times 10^{11}$~cm$^{-2}$.

Experimentally, the conductivity under such conditions has been studied in the greatest detail in wide quantum wells, $d = (70–80)$~nm, where the separation of states localized at the walls of the quantum well is strong, and the overlap of their wave functions is small \cite{Kvon08,Kvon11,Kozlov14,Zoth14,Dantscher15,Minkov19-1}. Typically, structures with a gate electrode are studied, which allows not only to change the carrier density by altering the position of the Fermi level, but also to change the symmetry of the quantum well. To analyze the results in this case, the simplest approximation is usually used, which assumes that the voltage $V_g$ applied to the gate electrode, while changing the energies of the electron branches in different ways, does not alter the charge distribution $e|\Psi(z)|^2$ in these branches ($e$ is the elementary charge, $\Psi$ is the wave function). In this approximation, a negative gate voltage at a certain value  reduces the concentration of electrons pressed against the wall of the quantum well closest to the gate to zero, so that with a further increase in the absolute value of the negative gate voltage, the conductivity should be determined by ``single-spin'' states pressed against the wall of the quantum well farthest from the gate and doubly degenerate holes. However, when studying electronic states under conditions where the Fermi level lies above the top of the valence band, we showed that a change in $V_g$ leads to a significant shift of the surface states in the $z$‑direction \cite{Minkov24}. Furthermore, as $V_g$ decreases, neither of the ``spin'' states disappears; both states persist up to $V_g$ values close to the charge neutrality point $V_\text{CNP}$. At the same time, the electron densities in the ``spin'' branches decrease at different rates as $V_g$ decreases, which predicts the disappearance of one of the spin states below the charge neutrality point, when the Fermi level drops below the valence band maximum, with a total electron concentration of approximately $0.5\times 10^{11}$~cm$^{-2}$
(for a quantum well with $d=32$~nm).

The aim of our study is to experimentally investigate electronic states under conditions where the Fermi level is located below the top of the valence band.

\label{sec:exp}

The studied structures Hg$_{1-x}$Cd$_x$Te/HgTe ($x=0.5–0.7$) were grown by molecular beam epitaxy on a (013)-GaAs substrate with a CdTe buffer layer $(5–7)$~$\mu$m thick \cite{Mikhailov06}. The measurements were carried out on Hall bridges with a channel width of $0.5$~mm and a distance of $0.5$~mm between the potential contacts, which were located $0.75$~mm away from the current contacts. The samples were mesa etched into standard Hall bars of $0.5$~mm width with the distance between the potential probes of $0.5$~mm. To change and control the electron and hole densities in the quantum well, the field-effect transistors were fabricated with parylene as an insulator and aluminum as a gate electrode. The measurements were performed in the \emph{dc}-regime of linear response at temperatures $1.4$~K in the magnetic
field up to $2.5$~T.

Structures with a quantum well width of $32$~nm were studied. The experimental results under conditions where the Fermi level lies above the top of the valence band and transport phenomena are determined solely by electrons were discussed in detail in \cite{Minkov24}.

\label{sec:res}

\begin{figure}[b]
\includegraphics[width=\linewidth,clip=true]{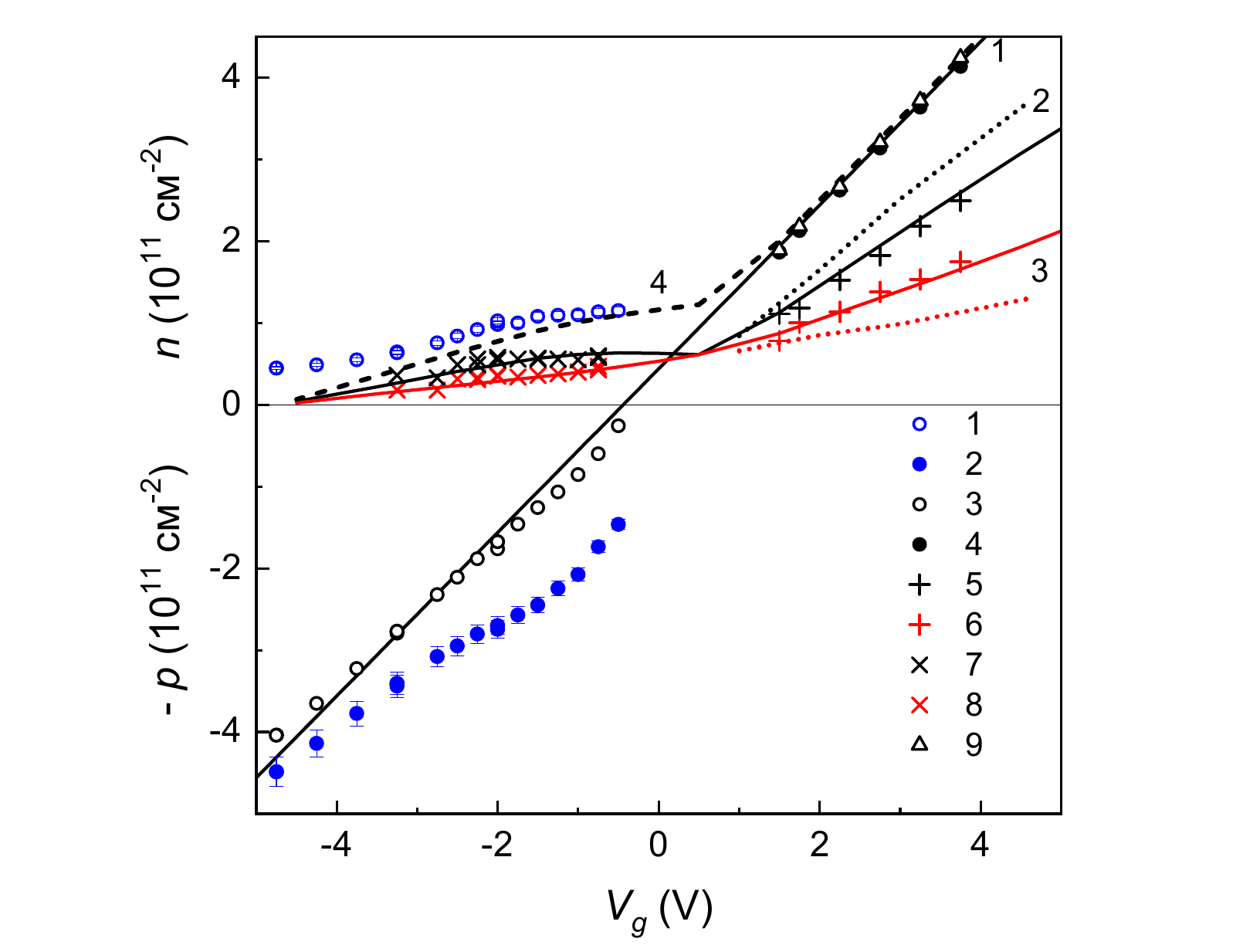}
\caption{(Color online) The $V_g$ dependences of the electron and hole densities obtained using various methods. Symbols $1$ and $2$ are the electron  and holes densities, respectively, obtained from the magnetic field dependences of $R_H$ and $R_{xx}$ in the model of conduction by two types of carriers; $3$ is the sum $n + p$; $4$ is the electron density obtained from the Hall effect as $-1/eR_H(0.3~\text{T})$; $5$ and $6$ are the electron density obtained from the Fourier spectra of the SdH oscillations at $V_g > 0$; $7$ and $8$ are the same as $5$ and $6$, but for $V_g <0$; $9$ are $n_1+n_2$. Line $1$ represents the dependence $(0.44+1.0 V_g, \text{V})\times 10^{11}$cm$^{-2}$, the slope of which equal $1.0$~V$^{-1}$ corresponds to the capacity measured experimentally; lines $2$ and $3$ represent the result of self‑consistent calculations with $\kappa_\text{QW} = 40$ (solid lines) and $20.5$ (dashed lines); line $4$ is the sum of lines $2$ and $3$.}
\label{f30}
\end{figure}

\begin{figure}
\includegraphics[width=\linewidth,clip=true]{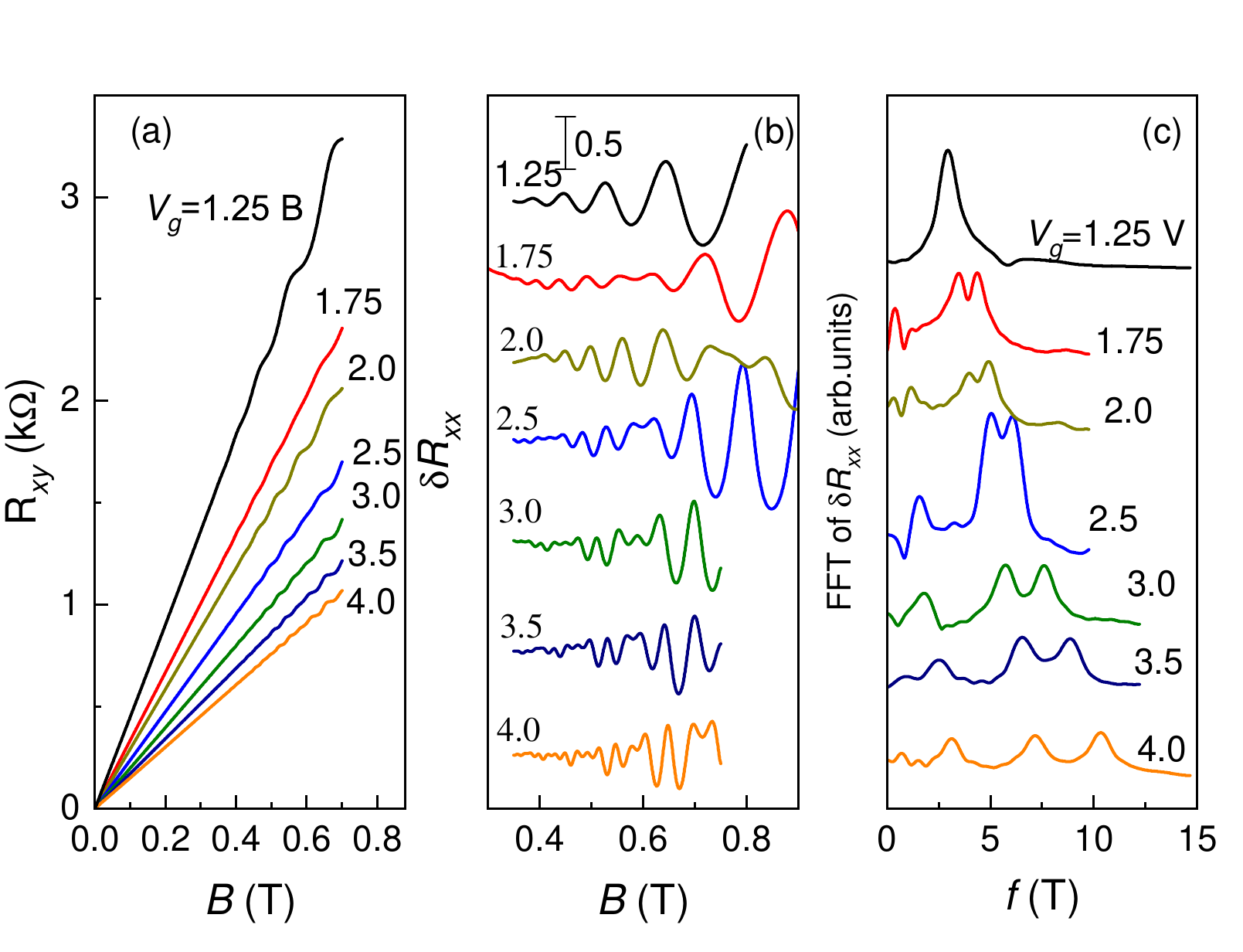}
\caption{(Color online) (a) and (b) -- The magnetic field dependences of $R_{xy}$ and  $\delta R_{xx}$, respectively. (c) The Fourier spectra of the oscillations shown the panel (b) for $V_g>0$.}
\label{f40}
\end{figure}

The $V_g$ dependences of the electron and hole densities are shown in Fig.~\ref{f30}. The electron concentration at $V_g > 1$~V was determined using two methods: from the Hall effect in the linear region of the $R_{xy}(B)$ variation and from the analysis of Shubnikov–de Haas (SdH) oscillations.  Fig.~\ref{f40} shows the magnetic field dependencies of  $R_{xy}$ and the oscillating part of $R_{xx}$:  $\delta R_{xx} = R_{xx}/R_\text{mon} – 1$, where $R_\text{mon}$ is the monotonic component of $R_{xx}(B)$. It is seen from Fig.~\ref{f40}(a) that the  dependence $R_{xy}(B)$ is linear  when $B < (0.3–0.4)$~T. Fig.~\ref{f40}(b) shows that above $V_g \simeq 1.5$~V, the oscillations develop a complex shape due to the onset of beats. Accordingly, two peaks emerge in the Fourier spectra [Fig.~\ref{f40}(c)], whose frequencies correspond to the electron concentrations $n_{1}$ and $n_{2}$ in the two spin–orbit-split branches.
As seen in Fig.~\ref{f30}, the sum $n_1 + n_2$ coincides with the electron density obtained from the Hall effect ($n_H$).

Let us compare these results with the results of a self‑consistent calculation consisting of simultaneous solving the Schr\"{o}dinger and  Poisson equations \cite{Minkov22}. The only parameter in this calculation was the dielectric constant in the quantum well $\kappa_\text{QW}$. Fig.~\ref{f30} shows the calculation results for two values of $\kappa_\text{QW}$ in the range of positive gate voltages. As  seen a good agreement with the experimental results is achieved at $\kappa_\text{QW}=40$, rather than at the value of $20.5$, which corresponds to bulk mercury telluride.

\begin{figure}
\includegraphics[width=\linewidth,clip=true]{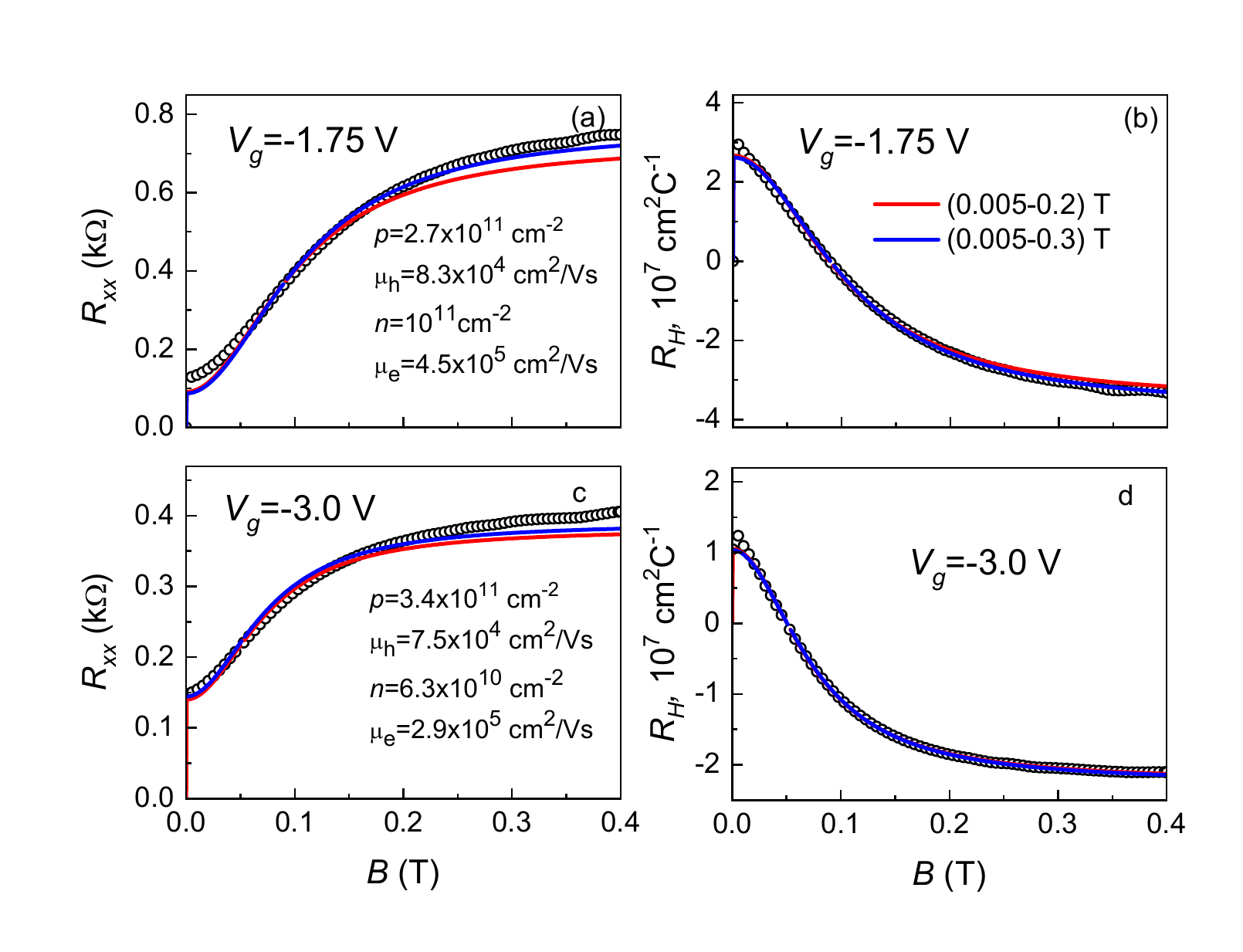}
\caption{(Color online) The dependencies $R_{xx}(B)$ (a), (c) and $R_H(B)$ (b), (d) for two $V_g$ values. The symbols represent the measurement results, and the lines represent the fitting results using a conductivity model with two types of carriers. Panels (a) and (c) show the parameters found as a result of the fitting. The fitting ranges are indicated in panel (b). }
\label{f50}
\end{figure}

We turn now to  the  negative $V_g$ values. In this $V_g$ range, both electrons and holes participate in conductivity. The electron density at $V_g < – 0.5$~V can be determined from a simultaneous analysis of the magnetic field dependences of $R_H$ and $R_{xx}$ using the standard two‑carrier model of conductivity. As an example, Fig.~\ref{f50} shows the experimental results (shown by symbols) and calculated curves corresponding to the best fit to the experiment in the range of magnetic fields from $0.005$~T to $0.2$~T and from $0.005$~T to $0.3$~T for two $V_g$ values: $V_g = -0.75$~V and $-3.0$~V. It is seen that the experimental results in both ranges are well described in this model, however, the error in determining the densities and mobilities of electrons and holes ($n$, $p$ and $\mu_e$, $\mu_h$, respectively) is quite large, about $(5-15)$~\%.

\begin{figure}
\includegraphics[width=0.9\linewidth,clip=true]{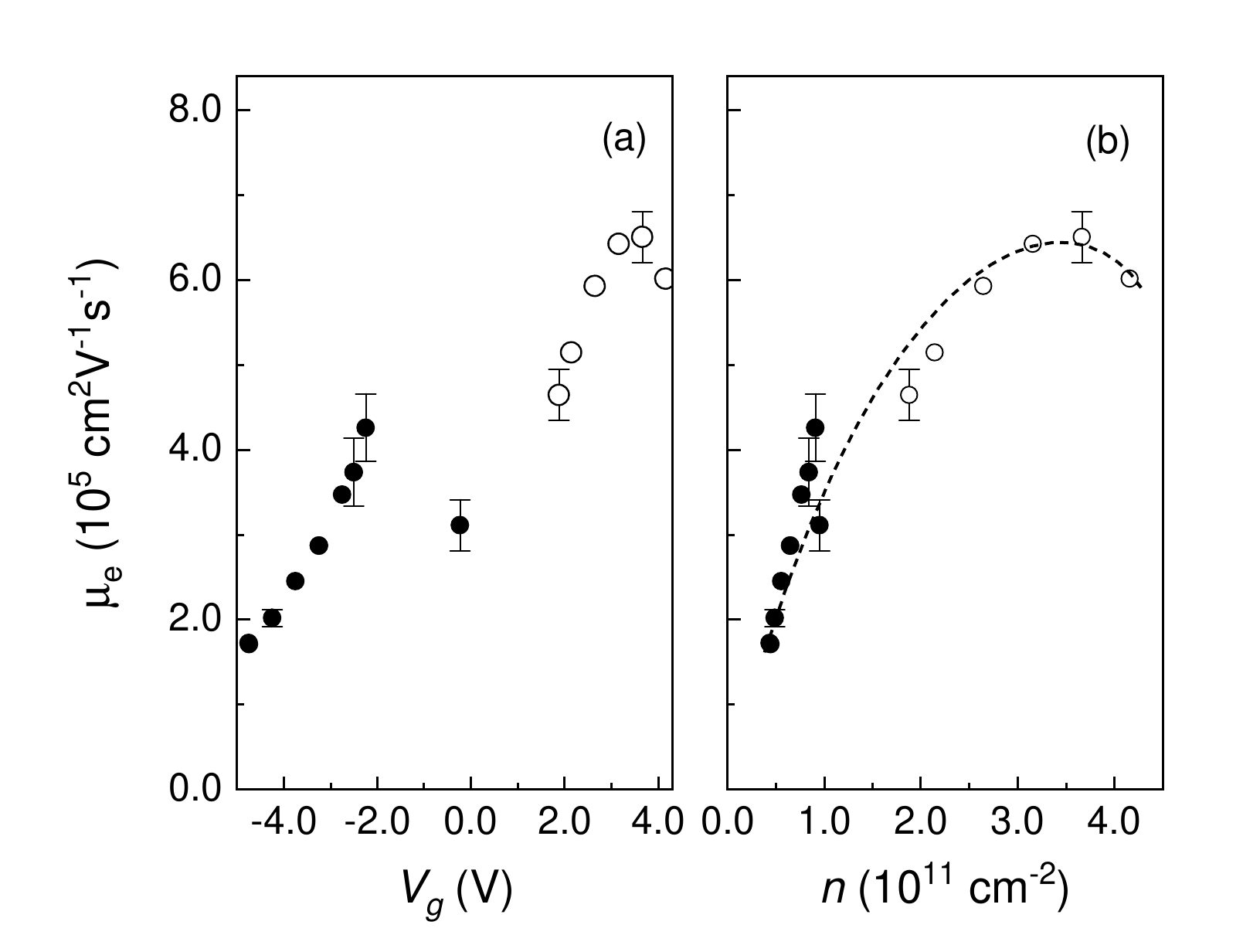}
\caption{(Color online) (a) and  (b) -- The $V_g$ an $n$ dependences of electron mobility, respectively. Solid symbols represent the results obtained using a two‑carrier conduction model; open symbols represent the Hall mobility $R_H(0.3~\text{T})/R_{xx}(0)$. The curve in panel (b) is a guide to the eye. }
\label{f60}
\end{figure}

The electron mobility, determined in the model of two types of carriers, depending on the voltage at the field electrode and the electron density, which is determined simultaneously with the mobility in this model, is shown in Fig.~\ref{f60}(a) and \ref{f60}(b), respectively. These figures also shows the Hall mobility obtained at $V_g > 0$, found as $R_H/R_{xx}$. It can be seen  that the mobility of electrons existing against the background of the valence band increases smoothly with increasing density, which is consistent with the electron mobility dependence in the positive gate voltage range.

As for the behaviour of the electron and hole densities, the gate voltage dependences of $n$ and $p$  follow a linear relation $(0.44 + 1.0\, V_g\text{ [V]} \times 10^{11})\text{cm}^{-2}$ (line $1$ in Fig.~\ref{f30}), whose slope agrees within experimental uncertainty with that derived from the 2D gas–gate capacitance of the same structure.

Despite the agreement between the experiment and the model, it should be noted that two approximations are made in the model calculations: 1) the conductivity is determined by two types of charge carriers; 2) all four parameters of this model -- $n$, $\mu_e$, $p$, and $\mu_h$ -- are independent of the magnetic field. The latter approximation may not hold well in cases of small band overlap; therefore, although the $n$ values obtained from fitting in different magnetic field ranges differ only slightly, the value of $n$ should be treated with caution.

\begin{figure}
\includegraphics[width=0.9\linewidth,clip=true]{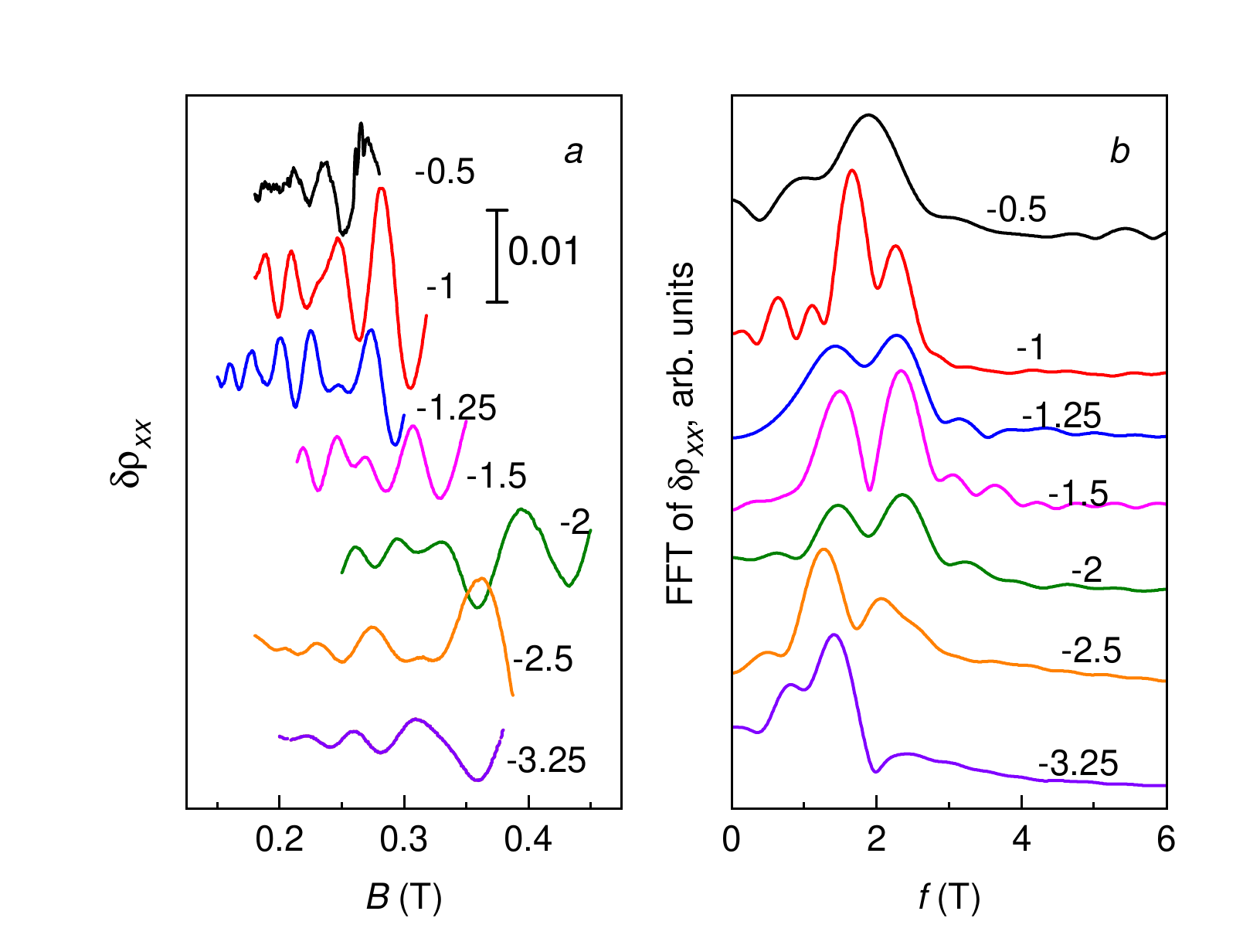}
\caption{(Color online) Oscillations $\delta R_{xx}(B)$ (a) and their Fourier spectra (b) for different $V_g$ values. }
\label{f70}
\end{figure}

\begin{figure}
\includegraphics[width=0.9\linewidth,clip=true]{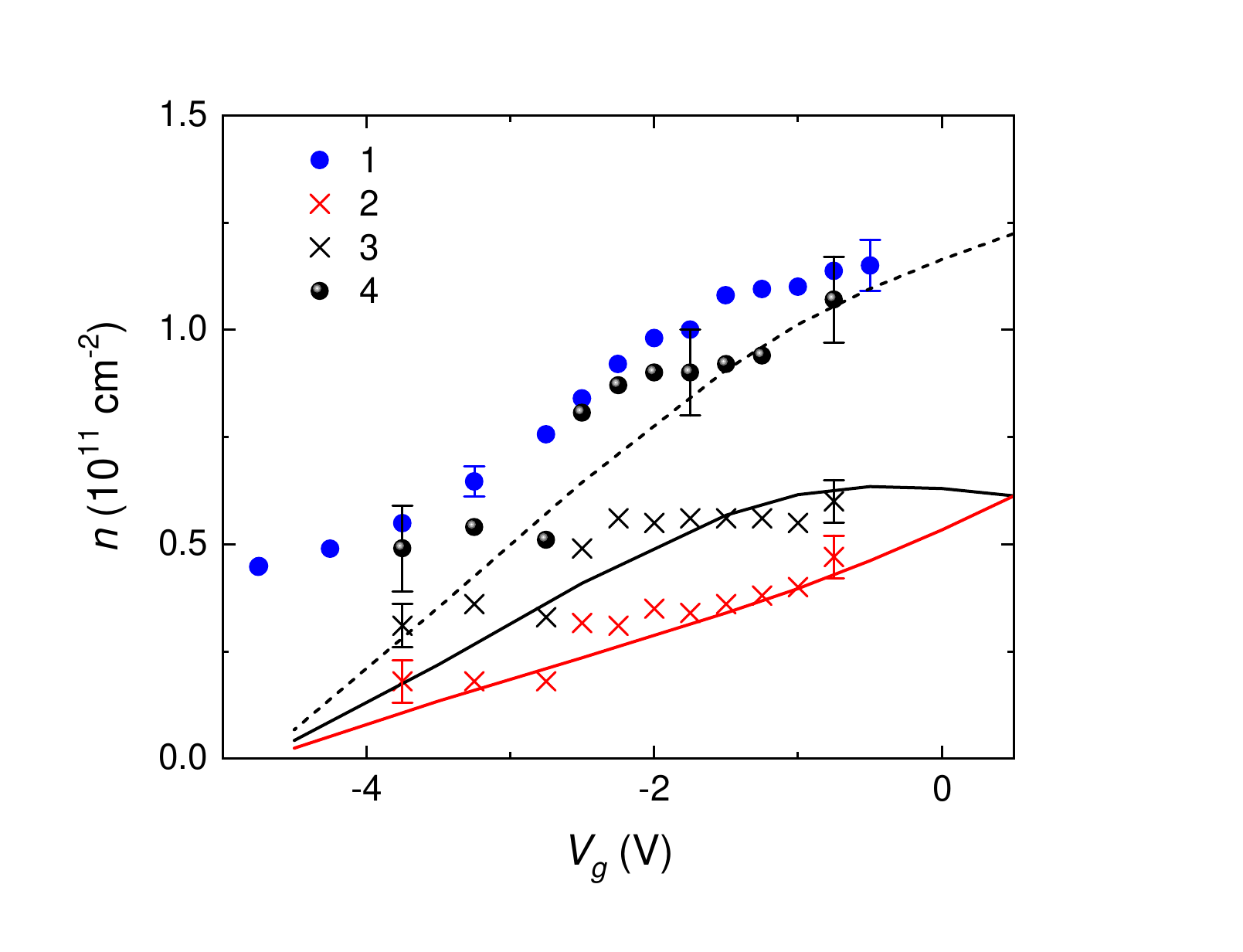}
\caption{(Color online) Electron densities plotted as a function of $V_g$ in the regime where electrons are against the background of the valence band. 1 -- the density  found in the conductivity model with two types of carriers; 2 and 3 -- the electron densities in the spin branches found from the Fourier analysis of the dependencies $\delta R_{xx}(B)$; 4 -- the sum of these concentrations. The black and red lines are the result of a self-consistent calculation of electron densities in the ``spin'' branches with  $\kappa_\text{QW}=40$, the dashed line is their sum.}
\label{f80}
\end{figure}

Similar results in wide HgTe quantum wells have been observed in numerous papers, where it was consistently assumed that the electron states coexisting with the valence band are ``single-spin'' states. From our point of view, the only experimental way to verify this claim is by comparing the electron densities derived from an analysis of the dependencies  $R_H(B)$ and $R_{xx}(B)$ with those determined from an analysis of SdH oscillations. To the best of our knowledge, oscillations originating from electronic states that coexist with valence-band states have not been reported previously. We believe this is because the SdH oscillations are not directly apparent in the $R_{xx}(B)$ data, due to the very sharp rise of $R_{xx}(B)$ at low magnetic fields. However, they become clearly discernible once the monotonic component is subtracted. Fig.~\ref{f70}(a) shows the  dependencies $\delta R_{xx}(B)$ for several $V_g$ values, while Fig.~\ref{f70}(b) presents the results of a Fourier analysis of these data. As seen  two peaks are clearly distinguishable in the Fourier spectra. The electron densities $n_1$ and $n_2$, corresponding to the frequencies of these peaks, are shown by diagonal crosses in Figs.~\ref{f30} and \ref{f80}.

Fig.~\ref{f80} shows that the electron density found in the conductivity model with two types of carriers within the error margin matches the sum of the electron concentrations in the two spin branches. A comparison of the experimental data with the results of self-consistent calculations using \(\kappa_\text{QW}=40\) shows good agreement both at negative values of \(V_{g}\), where electrons at the Fermi level coexist with valence-band states, and at positive values, where the Fermi level lies above the top of the valence band (see Figs.~\ref{f30} and \ref{f80}).

Possible reasons for the difference between the dielectric constant in the quantum well ($\kappa_\text{QW} = 40$) and that of the parent bulk material ($\kappa = 20.5$) were discussed in \cite{Andlauer09,Bruene14,Ziegler20}. One such reason may be the spatial quantization of the valence-band top. Indeed, the dielectric constant in semiconductors is mainly determined by the polarizability of the completely filled valence band. In some semiconductors (in particular, in HgTe), the  quantization of the valence-band top becomes significant even at relatively large quantum well widths, which can lead to a noticeable change in its polarizability.

\vspace{5mm}

In summary, the magnetotransport properties of $32$~nm-wide Hg$_{1-x}$Cd$_x$Te/HgTe quantum wells with a semimetallic spectrum were studied. A joint analysis of Shubnikov–de Haas oscillations and the magnetic-field dependences of the resistance and Hall coefficient at various gate voltages made it possible to determine the population of the spin branches of the electron spectrum split by the spin–orbit interaction. It was shown that, in contrast to the predictions of the idealized model, electrons populate both branches of the split spectrum regardless of whether the Fermi level is located solely in the conduction band or in the overlap region between the conduction band and the valence band. The obtained data are consistent with the energy spectrum calculations performed within a self-consistent four-band $kP$-model.

In summary, we studied the magnetotransport properties of $32$~nm-wide Hg$_{1-x}$Cd$_x$Te/HgTe quantum wells with a semimetallic spectrum. By combining the analysis of Shubnikov–de Haas oscillations with the magnetic-field dependences of the magnetoresistance and Hall coefficient across various gate voltages, we determined the carrier populations of the spin–orbit-split electron subbands. Contrary to the predictions of an idealized model, electrons populate both spin branches regardless of whether the Fermi level lies exclusively in the conduction band or within the conduction–valence band overlap region. These experimental findings agree well with self-consistent energy spectrum calculations based on the four-band $kP$-model.

\begin{acknowledgments}
The work has been supported  by  the Ministry of Science and Higher Education of the Russian Federation (Project No. FEUZ-2026-0014).
\end{acknowledgments}


\begin{thebibliography}{12}%
\makeatletter
\providecommand \@ifxundefined [1]{%
 \@ifx{#1\undefined}
}%
\providecommand \@ifnum [1]{%
 \ifnum #1\expandafter \@firstoftwo
 \else \expandafter \@secondoftwo
 \fi
}%
\providecommand \@ifx [1]{%
 \ifx #1\expandafter \@firstoftwo
 \else \expandafter \@secondoftwo
 \fi
}%
\providecommand \natexlab [1]{#1}%
\providecommand \enquote  [1]{``#1''}%
\providecommand \bibnamefont  [1]{#1}%
\providecommand \bibfnamefont [1]{#1}%
\providecommand \citenamefont [1]{#1}%
\providecommand \href@noop [0]{\@secondoftwo}%
\providecommand \href [0]{\begingroup \@sanitize@url \@href}%
\providecommand \@href[1]{\@@startlink{#1}\@@href}%
\providecommand \@@href[1]{\endgroup#1\@@endlink}%
\providecommand \@sanitize@url [0]{\catcode `\\12\catcode `\$12\catcode
  `\&12\catcode `\#12\catcode `\^12\catcode `\_12\catcode `\%12\relax}%
\providecommand \@@startlink[1]{}%
\providecommand \@@endlink[0]{}%
\providecommand \url  [0]{\begingroup\@sanitize@url \@url }%
\providecommand \@url [1]{\endgroup\@href {#1}{\urlprefix }}%
\providecommand \urlprefix  [0]{URL }%
\providecommand \Eprint [0]{\href }%
\providecommand \doibase [0]{https://doi.org/}%
\providecommand \selectlanguage [0]{\@gobble}%
\providecommand \bibinfo  [0]{\@secondoftwo}%
\providecommand \bibfield  [0]{\@secondoftwo}%
\providecommand \translation [1]{[#1]}%
\providecommand \BibitemOpen [0]{}%
\providecommand \bibitemStop [0]{}%
\providecommand \bibitemNoStop [0]{.\EOS\space}%
\providecommand \EOS [0]{\spacefactor3000\relax}%
\providecommand \BibitemShut  [1]{\csname bibitem#1\endcsname}%
\let\auto@bib@innerbib\@empty
\bibitem [{\citenamefont {Minkov}\ \emph {et~al.}(2022)\citenamefont {Minkov},
  \citenamefont {Aleshkin}, \citenamefont {Rut}, \citenamefont {Sherstobitov},
  \citenamefont {Dvoretski}, \citenamefont {Mikhailov},\ and\ \citenamefont
  {Germanenko}}]{Minkov22}%
  \BibitemOpen
  \bibfield  {author} {\bibinfo {author} {\bibfnamefont {G.~M.}\ \bibnamefont
  {Minkov}}, \bibinfo {author} {\bibfnamefont {V.~Y.}\ \bibnamefont
  {Aleshkin}}, \bibinfo {author} {\bibfnamefont {O.~E.}\ \bibnamefont {Rut}},
  \bibinfo {author} {\bibfnamefont {A.~A.}\ \bibnamefont {Sherstobitov}},
  \bibinfo {author} {\bibfnamefont {S.~A.}\ \bibnamefont {Dvoretski}}, \bibinfo
  {author} {\bibfnamefont {N.~N.}\ \bibnamefont {Mikhailov}},\ and\ \bibinfo
  {author} {\bibfnamefont {A.~V.}\ \bibnamefont {Germanenko}},\ }\bibfield
  {title} {\bibinfo {title} {Transformation of energy spectrum and wave
  functions on the way from a 2d-to-3d topological insulator in hgte quantum
  wells},\ }\href {https://doi.org/10.1103/PhysRevB.106.085301} {\bibfield
  {journal} {\bibinfo  {journal} {Phys. Rev. B}\ }\textbf {\bibinfo {volume}
  {106}},\ \bibinfo {pages} {085301} (\bibinfo {year} {2022})}\BibitemShut
  {NoStop}%
\bibitem [{\citenamefont {Kvon}\ \emph {et~al.}(2008)\citenamefont {Kvon},
  \citenamefont {Olshanetsky}, \citenamefont {Kozlov}, \citenamefont
  {Mikhailov},\ and\ \citenamefont {Dvoretskii}}]{Kvon08}%
  \BibitemOpen
  \bibfield  {author} {\bibinfo {author} {\bibfnamefont {Z.~D.}\ \bibnamefont
  {Kvon}}, \bibinfo {author} {\bibfnamefont {E.}~\bibnamefont {Olshanetsky}},
  \bibinfo {author} {\bibfnamefont {D.~A.}\ \bibnamefont {Kozlov}}, \bibinfo
  {author} {\bibfnamefont {N.~N.}\ \bibnamefont {Mikhailov}},\ and\ \bibinfo
  {author} {\bibfnamefont {S.~A.}\ \bibnamefont {Dvoretskii}},\ }\href@noop {}
  {\bibfield  {journal} {\bibinfo  {journal} {Pis'ma Zh. Eksp. Teor. Fiz.}\
  }\textbf {\bibinfo {volume} {87}},\ \bibinfo {pages} {588} (\bibinfo {year}
  {2008})},\ \translation{JETP Lett. \textbf{87}, 502 (2008)}\BibitemShut
  {NoStop}%
\bibitem [{\citenamefont {Kvon}\ \emph {et~al.}(2011)\citenamefont {Kvon},
  \citenamefont {Olshanetsky}, \citenamefont {Novik}, \citenamefont {Kozlov},
  \citenamefont {Mikhailov}, \citenamefont {Parm},\ and\ \citenamefont
  {Dvoretsky}}]{Kvon11}%
  \BibitemOpen
  \bibfield  {author} {\bibinfo {author} {\bibfnamefont {Z.~D.}\ \bibnamefont
  {Kvon}}, \bibinfo {author} {\bibfnamefont {E.~B.}\ \bibnamefont
  {Olshanetsky}}, \bibinfo {author} {\bibfnamefont {E.~G.}\ \bibnamefont
  {Novik}}, \bibinfo {author} {\bibfnamefont {D.~A.}\ \bibnamefont {Kozlov}},
  \bibinfo {author} {\bibfnamefont {N.~N.}\ \bibnamefont {Mikhailov}}, \bibinfo
  {author} {\bibfnamefont {I.~O.}\ \bibnamefont {Parm}},\ and\ \bibinfo
  {author} {\bibfnamefont {S.~A.}\ \bibnamefont {Dvoretsky}},\ }\bibfield
  {title} {\bibinfo {title} {Two-dimensional electron-hole system in hgte-based
  quantum wells with surface orientation (112)},\ }\href
  {https://doi.org/10.1103/PhysRevB.83.193304} {\bibfield  {journal} {\bibinfo
  {journal} {Phys. Rev. B}\ }\textbf {\bibinfo {volume} {83}},\ \bibinfo
  {pages} {193304} (\bibinfo {year} {2011})}\BibitemShut {NoStop}%
\bibitem [{\citenamefont {Kozlov}\ \emph {et~al.}(2014)\citenamefont {Kozlov},
  \citenamefont {Kvon}, \citenamefont {Olshanetsky}, \citenamefont {Mikhailov},
  \citenamefont {Dvoretsky},\ and\ \citenamefont {Weiss}}]{Kozlov14}%
  \BibitemOpen
  \bibfield  {author} {\bibinfo {author} {\bibfnamefont {D.~A.}\ \bibnamefont
  {Kozlov}}, \bibinfo {author} {\bibfnamefont {Z.~D.}\ \bibnamefont {Kvon}},
  \bibinfo {author} {\bibfnamefont {E.~B.}\ \bibnamefont {Olshanetsky}},
  \bibinfo {author} {\bibfnamefont {N.~N.}\ \bibnamefont {Mikhailov}}, \bibinfo
  {author} {\bibfnamefont {S.~A.}\ \bibnamefont {Dvoretsky}},\ and\ \bibinfo
  {author} {\bibfnamefont {D.}~\bibnamefont {Weiss}},\ }\bibfield  {title}
  {\bibinfo {title} {Transport properties of a 3d topological insulator based
  on a strained high-mobility hgte film},\ }\href
  {https://doi.org/10.1103/PhysRevLett.112.196801} {\bibfield  {journal}
  {\bibinfo  {journal} {Phys. Rev. Lett.}\ }\textbf {\bibinfo {volume} {112}},\
  \bibinfo {pages} {196801} (\bibinfo {year} {2014})}\BibitemShut {NoStop}%
\bibitem [{\citenamefont {Zoth}\ \emph {et~al.}(2014)\citenamefont {Zoth},
  \citenamefont {Olbrich}, \citenamefont {Vierling}, \citenamefont {Dantscher},
  \citenamefont {Bel'kov}, \citenamefont {Semina}, \citenamefont {Glazov},
  \citenamefont {Golub}, \citenamefont {Kozlov}, \citenamefont {Kvon},
  \citenamefont {Mikhailov}, \citenamefont {Dvoretsky},\ and\ \citenamefont
  {Ganichev}}]{Zoth14}%
  \BibitemOpen
  \bibfield  {author} {\bibinfo {author} {\bibfnamefont {C.}~\bibnamefont
  {Zoth}}, \bibinfo {author} {\bibfnamefont {P.}~\bibnamefont {Olbrich}},
  \bibinfo {author} {\bibfnamefont {P.}~\bibnamefont {Vierling}}, \bibinfo
  {author} {\bibfnamefont {K.-M.}\ \bibnamefont {Dantscher}}, \bibinfo {author}
  {\bibfnamefont {V.~V.}\ \bibnamefont {Bel'kov}}, \bibinfo {author}
  {\bibfnamefont {M.~A.}\ \bibnamefont {Semina}}, \bibinfo {author}
  {\bibfnamefont {M.~M.}\ \bibnamefont {Glazov}}, \bibinfo {author}
  {\bibfnamefont {L.~E.}\ \bibnamefont {Golub}}, \bibinfo {author}
  {\bibfnamefont {D.~A.}\ \bibnamefont {Kozlov}}, \bibinfo {author}
  {\bibfnamefont {Z.~D.}\ \bibnamefont {Kvon}}, \bibinfo {author}
  {\bibfnamefont {N.~N.}\ \bibnamefont {Mikhailov}}, \bibinfo {author}
  {\bibfnamefont {S.~A.}\ \bibnamefont {Dvoretsky}},\ and\ \bibinfo {author}
  {\bibfnamefont {S.~D.}\ \bibnamefont {Ganichev}},\ }\bibfield  {title}
  {\bibinfo {title} {Quantum oscillations of photocurrents in hgte quantum
  wells with dirac and parabolic dispersions},\ }\href
  {https://doi.org/10.1103/PhysRevB.90.205415} {\bibfield  {journal} {\bibinfo
  {journal} {Phys. Rev. B}\ }\textbf {\bibinfo {volume} {90}},\ \bibinfo
  {pages} {205415} (\bibinfo {year} {2014})}\BibitemShut {NoStop}%
\bibitem [{\citenamefont {Dantscher}\ \emph {et~al.}(2015)\citenamefont
  {Dantscher}, \citenamefont {Kozlov}, \citenamefont {Olbrich}, \citenamefont
  {Zoth}, \citenamefont {Faltermeier}, \citenamefont {Lindner}, \citenamefont
  {Budkin}, \citenamefont {Tarasenko}, \citenamefont {Bel'kov}, \citenamefont
  {Kvon}, \citenamefont {Mikhailov}, \citenamefont {Dvoretsky}, \citenamefont
  {Weiss}, \citenamefont {Jenichen},\ and\ \citenamefont
  {Ganichev}}]{Dantscher15}%
  \BibitemOpen
  \bibfield  {author} {\bibinfo {author} {\bibfnamefont {K.-M.}\ \bibnamefont
  {Dantscher}}, \bibinfo {author} {\bibfnamefont {D.~A.}\ \bibnamefont
  {Kozlov}}, \bibinfo {author} {\bibfnamefont {P.}~\bibnamefont {Olbrich}},
  \bibinfo {author} {\bibfnamefont {C.}~\bibnamefont {Zoth}}, \bibinfo {author}
  {\bibfnamefont {P.}~\bibnamefont {Faltermeier}}, \bibinfo {author}
  {\bibfnamefont {M.}~\bibnamefont {Lindner}}, \bibinfo {author} {\bibfnamefont
  {G.~V.}\ \bibnamefont {Budkin}}, \bibinfo {author} {\bibfnamefont {S.~A.}\
  \bibnamefont {Tarasenko}}, \bibinfo {author} {\bibfnamefont {V.~V.}\
  \bibnamefont {Bel'kov}}, \bibinfo {author} {\bibfnamefont {Z.~D.}\
  \bibnamefont {Kvon}}, \bibinfo {author} {\bibfnamefont {N.~N.}\ \bibnamefont
  {Mikhailov}}, \bibinfo {author} {\bibfnamefont {S.~A.}\ \bibnamefont
  {Dvoretsky}}, \bibinfo {author} {\bibfnamefont {D.}~\bibnamefont {Weiss}},
  \bibinfo {author} {\bibfnamefont {B.}~\bibnamefont {Jenichen}},\ and\
  \bibinfo {author} {\bibfnamefont {S.~D.}\ \bibnamefont {Ganichev}},\
  }\bibfield  {title} {\bibinfo {title} {Cyclotron-resonance-assisted
  photocurrents in surface states of a three-dimensional topological insulator
  based on a strained high-mobility hgte film},\ }\href
  {https://doi.org/10.1103/PhysRevB.92.165314} {\bibfield  {journal} {\bibinfo
  {journal} {Phys. Rev. B}\ }\textbf {\bibinfo {volume} {92}},\ \bibinfo
  {pages} {165314} (\bibinfo {year} {2015})}\BibitemShut {NoStop}%
\bibitem [{\citenamefont {Minkov}\ \emph {et~al.}(2019)\citenamefont {Minkov},
  \citenamefont {Aleshkin}, \citenamefont {Rut}, \citenamefont {Sherstobitov},
  \citenamefont {Germanenko}, \citenamefont {Dvoretski},\ and\ \citenamefont
  {Mikhailov}}]{Minkov19-1}%
  \BibitemOpen
  \bibfield  {author} {\bibinfo {author} {\bibfnamefont {G.}~\bibnamefont
  {Minkov}}, \bibinfo {author} {\bibfnamefont {V.~Y.}\ \bibnamefont
  {Aleshkin}}, \bibinfo {author} {\bibfnamefont {O.}~\bibnamefont {Rut}},
  \bibinfo {author} {\bibfnamefont {A.}~\bibnamefont {Sherstobitov}}, \bibinfo
  {author} {\bibfnamefont {A.}~\bibnamefont {Germanenko}}, \bibinfo {author}
  {\bibfnamefont {S.}~\bibnamefont {Dvoretski}},\ and\ \bibinfo {author}
  {\bibfnamefont {N.}~\bibnamefont {Mikhailov}},\ }\bibfield  {title} {\bibinfo
  {title} {Spin-orbit splitting of the conduction band in hgte quantum wells:
  Role of different mechanisms},\ }\href
  {https://doi.org/https://doi.org/10.1016/j.physe.2019.02.007} {\bibfield
  {journal} {\bibinfo  {journal} {Physica E: Low-dimensional Systems and
  Nanostructures}\ }\textbf {\bibinfo {volume} {110}},\ \bibinfo {pages} {95}
  (\bibinfo {year} {2019})}\BibitemShut {NoStop}%
\bibitem [{\citenamefont {Minkov}\ \emph {et~al.}(2024)\citenamefont {Minkov},
  \citenamefont {Rut}, \citenamefont {Sherstobitov}, \citenamefont {Dvoretski},
  \citenamefont {Mikhailov},\ and\ \citenamefont {Germanenko}}]{Minkov24}%
  \BibitemOpen
  \bibfield  {author} {\bibinfo {author} {\bibfnamefont {G.~M.}\ \bibnamefont
  {Minkov}}, \bibinfo {author} {\bibfnamefont {O.~E.}\ \bibnamefont {Rut}},
  \bibinfo {author} {\bibfnamefont {A.~A.}\ \bibnamefont {Sherstobitov}},
  \bibinfo {author} {\bibfnamefont {S.~A.}\ \bibnamefont {Dvoretski}}, \bibinfo
  {author} {\bibfnamefont {N.~N.}\ \bibnamefont {Mikhailov}},\ and\ \bibinfo
  {author} {\bibfnamefont {A.~V.}\ \bibnamefont {Germanenko}},\ }\bibfield
  {title} {\bibinfo {title} {Magnetotransport in hgte quantum wells with two
  gate electrodes},\ }\href {https://doi.org/10} {\bibfield  {journal}
  {\bibinfo  {journal} {Fiz. Tekn. Poluprov.}\ }\textbf {\bibinfo {volume}
  {58}},\ \bibinfo {pages} {434} (\bibinfo {year} {2024})}\BibitemShut
  {NoStop}%
\bibitem [{\citenamefont {Mikhailov}\ \emph {et~al.}(2006)\citenamefont
  {Mikhailov}, \citenamefont {Smirnov}, \citenamefont {Dvoretsky},
  \citenamefont {Sidorov}, \citenamefont {Shvets}, \citenamefont {Spesivtsev},\
  and\ \citenamefont {Rykhlitski}}]{Mikhailov06}%
  \BibitemOpen
  \bibfield  {author} {\bibinfo {author} {\bibfnamefont {N.~N.}\ \bibnamefont
  {Mikhailov}}, \bibinfo {author} {\bibfnamefont {R.~N.}\ \bibnamefont
  {Smirnov}}, \bibinfo {author} {\bibfnamefont {S.~A.}\ \bibnamefont
  {Dvoretsky}}, \bibinfo {author} {\bibfnamefont {Y.~G.}\ \bibnamefont
  {Sidorov}}, \bibinfo {author} {\bibfnamefont {V.~A.}\ \bibnamefont {Shvets}},
  \bibinfo {author} {\bibfnamefont {E.~V.}\ \bibnamefont {Spesivtsev}},\ and\
  \bibinfo {author} {\bibfnamefont {S.~V.}\ \bibnamefont {Rykhlitski}},\
  }\bibfield  {title} {\bibinfo {title} {Growth of hgcdte nanostructures by
  molecular beam epitaxy with ellipsometric control},\ }\href
  {https://doi.org/10.1504/IJNT.2006.008725} {\bibfield  {journal} {\bibinfo
  {journal} {Int. J. Nanotechnology}\ }\textbf {\bibinfo {volume} {3}},\
  \bibinfo {pages} {120 } (\bibinfo {year} {2006})}\BibitemShut {NoStop}%
\bibitem [{\citenamefont {Andlauer}\ and\ \citenamefont
  {Vogl}(2009)}]{Andlauer09}%
  \BibitemOpen
  \bibfield  {author} {\bibinfo {author} {\bibfnamefont {T.}~\bibnamefont
  {Andlauer}}\ and\ \bibinfo {author} {\bibfnamefont {P.}~\bibnamefont
  {Vogl}},\ }\bibfield  {title} {\bibinfo {title} {Full-band envelope-function
  approach for type-ii broken-gap superlattices},\ }\href
  {https://doi.org/10.1103/PhysRevB.80.035304} {\bibfield  {journal} {\bibinfo
  {journal} {Phys. Rev. B}\ }\textbf {\bibinfo {volume} {80}},\ \bibinfo
  {pages} {035304} (\bibinfo {year} {2009})}\BibitemShut {NoStop}%
\bibitem [{\citenamefont {Br\"une}\ \emph {et~al.}(2014)\citenamefont
  {Br\"une}, \citenamefont {Thienel}, \citenamefont {Stuiber}, \citenamefont
  {B\"ottcher}, \citenamefont {Buhmann}, \citenamefont {Novik}, \citenamefont
  {Liu}, \citenamefont {Hankiewicz},\ and\ \citenamefont
  {Molenkamp}}]{Bruene14}%
  \BibitemOpen
  \bibfield  {author} {\bibinfo {author} {\bibfnamefont {C.}~\bibnamefont
  {Br\"une}}, \bibinfo {author} {\bibfnamefont {C.}~\bibnamefont {Thienel}},
  \bibinfo {author} {\bibfnamefont {M.}~\bibnamefont {Stuiber}}, \bibinfo
  {author} {\bibfnamefont {J.}~\bibnamefont {B\"ottcher}}, \bibinfo {author}
  {\bibfnamefont {H.}~\bibnamefont {Buhmann}}, \bibinfo {author} {\bibfnamefont
  {E.~G.}\ \bibnamefont {Novik}}, \bibinfo {author} {\bibfnamefont {C.-X.}\
  \bibnamefont {Liu}}, \bibinfo {author} {\bibfnamefont {E.~M.}\ \bibnamefont
  {Hankiewicz}},\ and\ \bibinfo {author} {\bibfnamefont {L.~W.}\ \bibnamefont
  {Molenkamp}},\ }\bibfield  {title} {\bibinfo {title} {Dirac-screening
  stabilized surface-state transport in a topological insulator},\ }\href
  {https://doi.org/10.1103/PhysRevX.4.041045} {\bibfield  {journal} {\bibinfo
  {journal} {Phys. Rev. X}\ }\textbf {\bibinfo {volume} {4}},\ \bibinfo {pages}
  {041045} (\bibinfo {year} {2014})}\BibitemShut {NoStop}%
\bibitem [{\citenamefont {Ziegler}\ \emph {et~al.}(2020)\citenamefont
  {Ziegler}, \citenamefont {Kozlov}, \citenamefont {Mikhailov}, \citenamefont
  {Dvoretsky},\ and\ \citenamefont {Weiss}}]{Ziegler20}%
  \BibitemOpen
  \bibfield  {author} {\bibinfo {author} {\bibfnamefont {J.}~\bibnamefont
  {Ziegler}}, \bibinfo {author} {\bibfnamefont {D.~A.}\ \bibnamefont {Kozlov}},
  \bibinfo {author} {\bibfnamefont {N.~N.}\ \bibnamefont {Mikhailov}}, \bibinfo
  {author} {\bibfnamefont {S.}~\bibnamefont {Dvoretsky}},\ and\ \bibinfo
  {author} {\bibfnamefont {D.}~\bibnamefont {Weiss}},\ }\bibfield  {title}
  {\bibinfo {title} {Quantum hall effect and landau levels in the
  three-dimensional topological insulator hgte},\ }\href
  {https://doi.org/10.1103/PhysRevResearch.2.033003} {\bibfield  {journal}
  {\bibinfo  {journal} {Phys. Rev. Research}\ }\textbf {\bibinfo {volume}
  {2}},\ \bibinfo {pages} {033003} (\bibinfo {year} {2020})}\BibitemShut
  {NoStop}%
\end{thebibliography}

%

\end{document}